\documentclass[a4paper]{article}
\usepackage{amssymb}
\usepackage{pifont}
\newcommand{\cmark}{\ding{51}}%
\newcommand{\xmark}{\ding{55}}%

\usepackage{INTERSPEECH2020}

\title{Exploring Transformers for Large-Scale Speech Recognition}
\name{Liang Lu, Changliang Liu, Jinyu Li and Yifan Gong}
\address{Microsoft}
\email{\{liang.lu, chanliu, jinyli, yifan.gong\}@microsoft.com}

\begin{document}

\maketitle
\begin{abstract}
While recurrent neural networks still largely define state-of-the-art speech recognition systems, the Transformer network has been proven to be a competitive alternative, especially in the offline condition. Most studies with Transformers have been constrained in a relatively small scale setting, and some forms of data argumentation approaches are usually applied to combat the data sparsity issue. In this paper, we aim at understanding the behaviors of Transformers in the large-scale speech recognition setting, where we have used around 65,000 hours of training data. We investigated various aspects on scaling up Transformers, including model initialization, warmup training as well as different Layer Normalization strategies. In the streaming condition, we compared the widely used attention mask based future context lookahead approach to the Transformer-XL network. From our experiments, we show that Transformers can achieve around 6\% relative word error rate (WER) reduction compared to the BLSTM baseline in the offline fashion, while in the streaming fashion, Transformer-XL is comparable to LC-BLSTM with 800 millisecond latency constraint. 
\end{abstract}

\section{Introduction}

State-of-the-art speech recognition systems usually rely on the recurrent neural networks (RNNs) with long short-term memory (LSTM)~\cite{hochreiter1997long} units as their backbones. Recently, there have been an increasing interests in exploring Transformers~\cite{vaswani2017attention} for speech recognition, inspired by their success in nature language processing such as machine translation~\cite{vaswani2017attention} and language modeling~\cite{dai2018transformer}.  Compared to RNNs, Transformers do not process the input signal in a sequential fashion. Instead, they reply on the self-attention mechanism to capture the temporal correlations among the sequential signals, which circumvents the expensive back-propagation through time (BPTT)~\cite{werbos1990backpropagation} algorithm used to train RNNs. Therefore, Transformers can capture long-term correlations with much less computation complexity. Another advantage is that it is simpler to parallelize the computations in Transformers, which can reduce the time to train deeper models on a large scale dataset.  

For speech recognition, Transformers have achieved competitive recognition accuracy compared to RNN-based counterparts within both end-to-end~\cite{karita2019comparative, dong2018speech, sperber2018self, tian2019self, salazar2019self, zeyer2019comparison} and hybrid~\cite{han2019state, wang2019transformer} frameworks. However, the superior results are usually achieved in the offline condition, while in the streaming fashion, Transformers have shown significant degradation in terms of accuracy from previous results~\cite{karita2019comparative, wang2019transformer}, even in a condition of a large latency constraint. For example, \cite{wang2019transformer} reports around 25\% - 40\% accuracy degradation compared with the offline baseline with approximately 2.5 seconds of latency when using a hybrid model, while in~\cite{moritz2020streaming}, the streaming Transformer with around 1.2 second latency lagged behind its offline baseline by approximately 15\% - 23\% from experiments with an sequence-to-sequence (S2S) model. In addition, the studies of Transformer model mostly focus on a relatively small dataset from a single domain such as Librispeech dataset, and some forms of data argumentation approaches are applied to tackle the data sparsity issue. It is relatively less well understood where Transformer stands in a large-scale setting. 

In this paper, we aim at understanding the behaviors of Transformers for large scale speech recognition, and performing a fair comparison to LSTMs in both offline and streaming conditions. We use around 65,000 hours of training data, and apply data-parallelization across up to 64 GPUs.  We present our approaches to address the technical challenges in cross-machine multi-GPU training of deep Transformers, including model initialization and warmup training. In the offline condition, we show that our Transformers can outperform the bi-directional LSTM (BLSTM) \cite{graves2013hybrid} baseline with approximately 6\% relative with similar number of model parameters, while in the streaming fashion, we introduce the Transformer-XL~\cite{dai2018transformer} based steaming model, which is computationally tractable for inference. Our results show that Transformer-XL is on par with latency-controlled BLSTM (LC-BLSTM) \cite{HighwayBLSTM-zhang2016} with the same latency constraint. 


\section{Related Work}

There have been a few studies on Transformers for end-to-end speech recognition, particularly in the context of the S2S model with attention~\cite{karita2019comparative, dong2018speech, sperber2018self, wang2019semantic}, as well as Transformer Transducers~\cite{tian2019self, zhang2020transformer}.  In~\cite{karita2019comparative, zeyer2019comparison}, the authors compared RNNs with transformers for various speech recognition tasks, and obtained competitive or even better results with Transformers. Within the hybrid framework, the authors in~\cite{han2019state, wang2019transformer} also reported very strong results on the Librispeech benchmark with Transformers. The time restricted self-attention investigated in~\cite{povey2018time} is closely related to the Transformer-XL network in this work as they both perform chunk-wise training. The key difference is that in Transfromer-XL, we also take the hidden states from previous chunk as feature to capture the long-term information from the past. In terms of streamable Transformers, attention mask based approach to control the left and right context is mostly studied in the previous works, such as in the S2S model~\cite{moritz2020streaming}, Transformer-Transducer~\cite{zhang2020transformer} and the hybrid model~\cite{wang2019transformer}. From the reported results, streaming Transformers still lag behind their offline counterparts remarkably in terms of the recognition accuracy, especially in the low-latency scenarios. 

\section{Transformer Model}
\label{sec:transformer}


\subsection{Self-attention with Multiple Heads}

The attention mechanism is well understood~\cite{bahdanau2014neural}. In~\cite{vaswani2017attention}, the authors introduced a machine translation model based on self-attention, which can be represented in the form of the dot-product attention as: 
\begin{align}
\label{eq:att}
\text{Attention}(Q, K, V) = \text{Softmax}\left(\frac{QK^T}{\sqrt{d_k}}\right)V, 
\end{align}
where $Q,K,V$ are referred to the query, key and value according to~\cite{vaswani2017attention}. In self-attention, $Q, K$ and $V$ are from the source sequence, while in the conventional attention model~\cite{bahdanau2014neural}, $Q$ is from the decoder hidden state, and $K$ is from the encoder hidden state. In Eq \eqref{eq:att}, $d_k$ is the dimension of the model. 

Another key idea in~\cite{vaswani2017attention} is the multi-head attention mechanism, which performs multiple attention operations in parallel using different model parameters. The outputs from different attention heads are then concatenated and projected before being fed into the next layer, which can be shown as:
\begin{align}
\text{MultiHead}(Q, K, V) &= [H_1, H_2, \cdots, H_N]W^O \\ \nonumber
\text{where } H_i & = \text{Attention}(QW_i^Q, KW_i^K, VW_i^V) 
\end{align}
where $N$ is the number of attention heads, and $W_i^Q, W_i^K, W_i^V$ are parameters for the $i$-th attention head, and $W^O$ is the projection matrix to reduce the dimension of the concatenated vector.

\subsection{Depth-scale Initialization and Warmup Training}
\label{sec:init}

Training neural networks in a large scale setting requires parallelization across multiple GPUs. In our experiments, we did not experience convergence issues when training a shallow Transformer with random initialization on a single machine with 4-8 GPUs. However, we observed poor convergence or even divergence when performed parallelization across multiple machines with 32-64 GPUs with a randomly initialized model. To address this problem, we performed warmup training on a single machine with 4 GPUs until the model has seen $\sim$640 hours of training data, before switching to cross-machine parallelization. This approach worked well for Transformers with up to 12 layers in our experiments, however, when we increased the depth of the Transformers to 24 layers and beyond, the model did not converge even during the warmup stage. This is primarily due to the gradient explosion and vanishing problem in deep Transformers as investigated in~\cite{zhang2019improving}, which shows the gradient norms of lower layers were significantly smaller that those of deeper layers. This indicates that the supervision signal becomes much weaker after back-propagation from the top layer to lower layers. To address this problem, the authors proposed the depth-scale model initialization, which normalized the $\ell_2$-norm of the weight parameters by their corresponding depth index. For the model parameters from the $l$-th layer, they can be initialized as: 
\begin{align}
 W \in \mathbb R^{d_i \times d_o} \sim \mathcal{U}\left(-\frac{\gamma}{\sqrt{l}}, \frac{\gamma}{\sqrt{l}}\right), \gamma = \sqrt{\frac{6}{d_i + d_o}},
\end{align}
where $d_i$ and $d_o$ are input and output dimension respectively. 
With depth-scale initialization, we did not observe convergence issues during warmup training, and were able to train a Transformer-XL with 100 layers and over 200 million parameters in the warmup stage. However, we did not train this model until convergence due to the high computational cost.

\subsection{Pre-Norm vs. Post-Norm}
\label{sec:norm}

Layer normalization (LN)~\cite{ba2016layer} has been a {\it de facto} in Transformers for a smooth model convergence during training. In the seminal Transformer paper~\cite{vaswani2017attention}, LN is applied after the element-wise residual addition, i.e., 
\begin{align}
x_{l+1} = \text{LN}(x_l + \mathcal{F}(x_l, \theta_l)),
\end{align}
where $\theta_l$ denotes all the model parameters in the $l$-th layer, and $\mathcal{F}(\cdot)$ denotes any type of non-linear operation. This normalization approach, referred to as Post-Norm, is observed to results in poor convergence in machine translation tasks for deeper Transformers~\cite{wang2019learning}. In our experiments, the Post-Norm approach worked well for 12-layer Transformers trained with data-parallelization on a single machine with 4-8 GPUs. However, when we run cross-machine parallelization with 32 or 64 GPUs, we observed divergence during training with Post-Norm. As investigated in~\cite{wang2019learning}, Post-Norm poses a high risk of vanishing or exploding gradients problem, which is more apparent in the multi-GPU training with a very large minibatch size. We tackled this problem by switching to Pre-Norm as proposed in~\cite{wang2019learning}, where the LN is applied as
\begin{align}
x_{l+1} = x_l + \mathcal{F}(\text{LN}(x_l), \theta_l).
\end{align}
The Pre-Norm approach was also applied in~\cite{wang2019transformer}. With this approach, we were able to train a 48-layer Transformer-XL model with 64 GPUs without any convergence issues. 

\begin{figure}[t]
\small
\centerline{\includegraphics[width=0.32\textwidth]{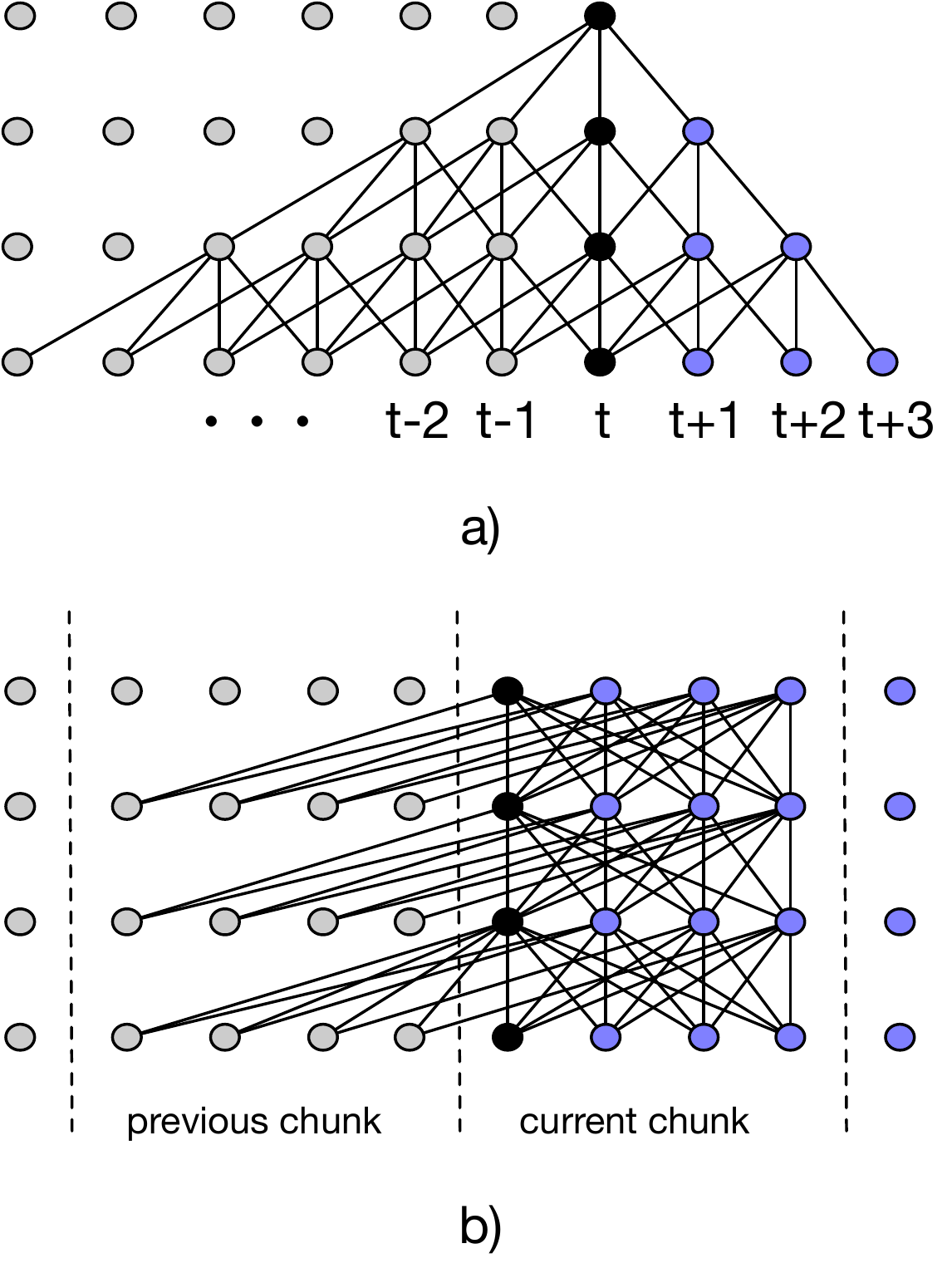}}
\vskip-4mm
\caption{a) Context expansion for a streaming Transformer based on attention masks. In this example, the attention context window is $[-2, 1]$ for each layer. With 3-layers, the context size increases to $[-6, 3]$. The gray nodes denote the past frames and hidden states, and the blue nodes represent the future frames and hidden states. b) A Transformer-XL with chunk size as 4. The context size does not changes as the model goes deeper.}  
\label{fig:mask}
\vskip -5mm
\end{figure}

\subsection{Streaming Transformer}
\label{sec:stream}

ASR systems are usually deployed in the streaming fashion which produces real-time speech transcription with certain latency constraint. To enable Transformers for streaming ASR, we have to limit the feature context visible to the model in both training and inference stage. A typical approach is based on the future context lookahead with an attention mask as shown in Figure~\ref{fig:mask}.a), where each node can only access limited number of activation states from both the future and the past timesteps. However, this approach suffers from the issue of context expansion, as the context size increases as the model goes deeper. Consequently each layer can only access features from a very small context window in order to fit into the latency constraint. This hinders the model to unlock the information from the data, and possibly explain the inferior recognition accuracy observed in this work as well as in~\cite{wang2019transformer, moritz2020streaming} when employing this approach.  Another disadvantage of this method is the high computational cost, as all the nodes in the context window need to be updated for every decoding step. This results in intractable computational complexity for real-time inference. 

In this work, we explore another streaming approach base on the Transformer-XL network~\cite{dai2018transformer}. Unlike the mask-based lookahead approach, Transformer-XL consumes the input sequence in a chunk-wise fashion, which means the information is fully visible to all the nodes when they reside in the same chunk. In order to capture the information from the long-term history,  Transformer-XL takes the hidden states from the previous chunk as additional feature, while does not pass the gradients through those hidden states to avoid the memory explosion. More precisely, this operation can be represented as:
\begin{align}
&\tilde{H}_{c-1}^l = {\text SG}(H_{c-1}^l), \\
&\tilde{K}_c^l = \text{Concat}[\tilde{H}_{c-1}^l; K_c^l],  \tilde{V}_c^l = \text{Concat}[\tilde{H}_{c-1}^l; V_c^l], \\
&H_c^l = \text{MultiHead} (Q_c^l, \tilde{K}_c^l, \tilde{V}_c^l),
\end{align}
where $c$ denote the index of the chunk, SG refers to the stop-gradient operation, and Concat indicates the concatenation operation of the two matrices along the time axis. The interpretation is that in the self-attention layer, the queries are from the current chunk, while the keys and values are from the current chunk as well as the previous chunk.

Transformer-XL enjoys two key advantages compared with the mask-based lookahead approach for streaming ASR. Firstly, we can emit the labels of all the acoustic frames within the chunk without refreshing the hidden states of self-attention, and therefore significantly cut down the computational cost for inference. Secondly, Transformer-XL enables us to train much deeper models. Recall that the self-attention operation has the memory complexity of $O(T^2)$, where $T$ is length of the input sequence. In Transformer-XL, the length of the chunk can be much shorter, which saves memory for deeper or larger models. As aforementioned, we were able to train a Transformer-XL with 100 layers without blowing up the GPU memory in the warmup stage. However, Transformer-XL has the disadvantage of longer training time due to the chunk-wise processing.

\section{Experiments and Results}
\label{sec:exp}

\subsection{Experimental setup}

In our experiments, the models were trained with around 65,000 hours of anonymized and transcribed Microsoft  data, recorded in various conditions. We fit the Transformer models into the hybrid architecture, where we used the hidden Markov models (HMMs) for sequence modeling. The number of tied triphone states is 9404 in our experiments, and the input feature is 80-dimension log Mel filter banks sampled every 20 milliseconds (ms) with the frame skipping strategy \cite{Miao16}. The language model is a 5-gram with around 100 million (M) ngrams. We used a {\tt dev} set with around 7459 utterances, which are spontaneous conversational speech recorded in both close-talk and far-filed conditions mostly from non-native speakers. The speaking style and acoustic condition of the {\tt dev} set are not well represented by our training data, so that we can avoid over-training by tuning on this dataset. Our {\tt eval} set is from the far-field scenario mostly by native speakers, which has around 13,368 utterances, and it matches our training data well.  

We compare Transformers with a bidirectional LSTM (BLSTM) based acoustic model in the offline condition, and an LC-BLSTM~\cite{HighwayBLSTM-zhang2016} in the streaming condition. Both BLSTM and LC-BLSTM in our experiments have 6 hidden layers, and each layer has 600 hidden units and cells for each direction. For LC-BLSTM, the chunk size is set to be 40. For Transformers, to limit the scope of our investigate, we set the dropout ratio to be 0.1, and the number of hidden units in the feedforward layer to be 2048 in all our experiments. All the models were trained with the cross-entropy criterion. 
We used the Adam optimizer~\cite{kingma2014adam} to train all our Transformer and LSTM models. For 12-layer Transformers, we set the mini-batch size as 16,000 frames for each GPU, and it took around 4 days for the models to converge with 32 Tesla V100 GPUs, which corresponds to training the models for 5-6 epochs. The training time for deeper Transformers is longer. For Transformers with 24 layers and beyond, we halved the mini-batch size due to the memory constraint, and tuned the learning rate scheduler accordingly. This increased the training time by roughly 50\%. Transformer-XL is much more expensive due to the chunk-wise training fashion. Its training time is approximately 50\% longer compared with a vanilla Transformer with the same model structure. For a 48-layer Transformer-XL, the training time is around 3 weeks with data parallelization across 64 Tesla V100 GPUs.

\begin{table}[t]\centering
\caption{Results of the Transformers with convolution layers in the offline mode. The number of heads is 4, and the number of layers is 12. {\tt IC} stands for interleaved 1D convolution. All models have around 50 million (M) model parameters including convolution layers. }
\label{tab:conv}
\footnotesize
\vskip-2mm
\begin{tabular}{l|lcccc}
\hline 

\hline
Model & {\tt IC}          & Size(M) & Encode layer & $d_k$  & {\tt dev} \\ \hline
  &\xmark & 51.5 & Linear & 620 & 34.7 \\ 
Transformer &\cmark & 50.0 & Linear & 512&  20.2 \\ 
&\cmark & 51.5  & VGG & 512 & 19.6 \\
&\xmark & 52.0  & VGG & 620 & 19.4 \\

\hline

\hline
\end{tabular}
\vskip-3mm
\end{table}

\begin{table}[t]\centering
\caption{Results of the 12-layer Transformer model with different number of attention heads. VGG net was used as the encoding layer for all the Transformers. $N$ denotes the number of attention heads, and $d_k$ is the model dimension as in Eq(\ref{eq:att}).}
\label{tab:head}
\footnotesize
\vskip-2mm
\begin{tabular}{l|lcccc}
\hline 

\hline
Model & {\tt IC}          & Size (M) & $N$ & $d_k$  & {\tt dev}  \\ \hline
& \cmark & 50.5  & 4 & 512 & 19.6 \\
& \cmark & 50.5  & 8 & 512 & 19.7  \\
& \cmark & 50.5 & 16 & 512 & 18.8  \\
Transformer &\xmark & 52.0 & 4 & 620 & 19.4  \\
& \xmark & 53.5 & 8 & 624 & 18.4  \\
& \xmark & 53.5 & 16 & 624 &  18.6 \\ \hline
BLSTM & -- & 55.0 & -- & --  & 19.5 \\
\hline

\hline
\end{tabular}
\vskip-5mm
\end{table}

\subsection{Convolution Layers and Attention Heads}

The self-attention operation cannot maintain the monotonicity of input sequence, which is particularly harmful for time-synchronous acoustic model such as the hybrid model studies in this paper. The positional encoding approach in~\cite{vaswani2017attention} is shown to be less effective to the speech recognition problem~\cite{wang2019transformer, lu2019transformer}, while convolutional layers are proven to be more powerful to encode the positional information. In Table~\ref{tab:conv}, we compare the two schemes of using convolution layers in Transformers in the offline condition, namely, the interleaved 1D convolution with self-attention from our previous study~\cite{lu2019transformer}, and using the VGG net~\cite{simonyan2014very} as the input encoding layer. The kernel size for the 1D convolution is 3, while the VGG net has 4 layers of 2D convolutions with 3x3 filters. When the VGG encoder was applied, we used features of frame rate as 10ms, and employed a max-pooling layer to down-sample the features by a factor of 2. As shown in Table~\ref{tab:conv}, the Transformer model performed poorly without any convolution layers. Both interleaved convolution and VGG net can significant improve the accuracy of Transformer models. In addition, when applying a VGG as the encoder, it is more beneficial to remove the interleaved convolutions but increase the model dimension of self-attention layers if the model size is constrained to be the same. 

Table~\ref{tab:head} shows the results of the Transformers with different numbers of attention heads. With the interleaved convolution, the Transformer with 16 attention heads achieved the lowest WER, while for the vanilla Transformer, 8 attention heads are sufficient. We did not further increase the number of the attention heads in our experiments due to the memory constraint. Compared to the BLSTM with around 55 million model parameters, the Transformer can achieve around 6\% relative WER reduction with a similar model size. 

\begin{table}[t]\centering
\caption{Results of streaming Transformer models. The number of layers is 12. }
\label{tab:stream}
\scriptsize
\vskip-2mm
\begin{tabular}{l|l|cccccc}
\hline 

\hline
Model & {\tt IC}          & Size (M) & $N$ & $d_k$ & Context  & {\tt dev}      \\ \hline
& \cmark & 50.5 & 16 & 512 &  [-$\infty, \infty$] & 18.8  \\
& \cmark & 50.5 & 16 & 512 & [-$\infty, 16$] & 20.6  \\
 & \cmark &50.5 & 16 & 512 & [-$\infty, 28$] & 20.7  \\
& \cmark & 50.5 & 16 & 512 & [-$\infty, 40$] & 20.0  \\ \cline{2-7}
Transformer& \xmark & 53.5 & 8 & 624 & [-$\infty, \infty$] & 18.4  \\
& \xmark & 53.5 & 8 & 624 & [-$\infty, 4$] & 23.0 \\
& \xmark & 53.5 & 8 & 624 & [-$\infty, 16$] & 21.1    \\
& \xmark & 53.5 & 8 & 624 & [-$\infty, 28$] & 21.8    \\
& \xmark & 53.5 & 8 & 624 &  [-$\infty, 40$] & 19.8    \\ \hline
Transformer-XL & \cmark & 50.5 & 16 & 512 & [-40, 40] & 20.4 \\
& \xmark & 53.5 & 8 & 624 & [-40, 40] & 21.0 \\ \hline
BLSTM & -- & 55.0 & -- & -- & $[-\infty, \infty]$ & 19.5 \\
LC-BLSTM & -- & 55.0 & -- &  -- & [-1, 40] & 20.2 \\
\hline
\end{tabular}
\vskip-5mm
\end{table}

\subsection{Results of Streaming Transformers}

The previous experiments focused on the offline scenario. In this section, we evaluate the accuracy of Transformers in the steaming condition. We investigated the mask-based lookahead approach, as well as the Transformer-XL network discussed in section~\ref{sec:stream}. The results are shown in Table~\ref{tab:stream} with various latency constraints. For Transformers with attention masks, the context window in Table~\ref{tab:stream} refers to the overall accumulated latency from both convolution and self-attention. For instance, $[-\infty, 40]$ corresponds to looking ahead 3 frames for each self-attention layer in a 12-layer Transformer without interleaved convolution, with an additional 4 frames latency from the VGG encoder. Since our Transformers operated at the 20 ms frame rate, 40 frames correspond to 800 ms latency. For the attention mask based Transformer, we did not limit the left context, so it is marked as $-\infty$, while the context window as $[-\infty, \infty]$ refers to the offline system. For Transformer-XL, we set the chunk size as 40, and since the model takes the hidden states from the previous chunk as feature, we denote the context size as $[-40, 40]$.  Note that there are not overlaps in the chunks during both training and inference, and it emits 40 outputs each time during inference. For LC-BLSTM, the chunk size is also 40, and because it takes the previous hidden state as the representation of history, arguably, we denote the context size as $[-1, 40]$. The chunks in LC-BLSTM are overlapped by 20 frames, so it only emits 20 outputs each time during inference. 

From the results in Table~\ref{tab:stream}, we have the following observations. First, without the interleaved convolution, the streaming Transformers based on attention masks degrade the recognition accuracy by 18\% - 25\% compared to the offline baseline model in the low latency condition. With the interleaved convolution, the accuracy loss is much smaller. This may be due to that the interleaved convolution layers can compensate the reordering effect of self-attention operations, and maintain the monotonicity of the input sequence~\cite{lu2019transformer}. When the latency constraint is less tight, the effect of interleaved convolution layers is diminishing. Second, Transformer-XL still lags behind the vanilla Transformer with the context window of $[-\infty, 40]$. This is not surprising as in Transformer-XL, the previous chunk is only an approximation of the full history. Improving the strength of Transformer-XL to capture the information from the long-term history in a computationally feasible manner, such as the Compressive Transformer~\cite{rae2019compressive}, is worth further investigation. Third, the gap between the streaming Transformer (or Transformer-XL) and its offline model is larger than that between LC-BLSTM and BLSTM. The offline Transformer outperforms BLSTM by a considerable margin, while Transformer-XL is only comparable with LC-BLSTM in terms of WERs. Though the Transformer with attention mask as $[-\infty, 40]$ can outperform LC-BLSTM, it is not computationally feasible during inference. This observation may align well with the argument that Transformers are more powerful to capture the long-term correlations in sequential signals. In a scenario with limited feature context, however, Transformers are hindered to release their modeling power. 

\subsection{Deeper Transformers}

In Table~\ref{tab:deeper}, we show results from deeper Transformers, and results from the {\tt eval} set. We observe similar trend on the {\tt eval} set. We also investigated the tradeoff between increasing the dimension of hidden state in self-attention layers and increasing the depth the model. For the offline 12-layer Transformer, we increased $d_k$ to 960, which resulted in a model with 97 million parameters. However, we only achieved very marginal improvement on the {\tt dev} set. The gain from increasing the model depth to 24 layers is more considerable on the {\tt dev} set, but the gain on the {\tt eval} set is still small. It is possible that the model is overfitted, and increasing the dropout ratio may result in more gains. As for Transformer-XL, we can obtain gains in terms of accuracy by increasing the depth up to 48 layers. However, the gains are not as large as we have expected, and regularizing the deep Transformers may result in further improvements.  

\begin{table}[t]\centering
\caption{Results of deeper Transformer models. $L$ denotes the model depth.}
\label{tab:deeper}
\scriptsize
\vskip-2mm
\begin{tabular}{l|l|cccccc}
\hline 

\hline
Model & {\tt IC}          & Size(M) & $L$ & Context  & {\tt dev} & {\tt eval}      \\ \hline
BLSTM & -- & 55.0  & 6 &  [-$\infty, \infty$] &  19.5 & 12.7  \\
LC-BLSTM & -- & 55.0  & 6 &  [-1, 40] &  20.2 & 12.9  \\ \hline

 & \xmark & 53.5  & 12 &  [-$\infty, \infty$] & 18.4  & 11.9  \\
Transformer & \xmark & 97.0  & 12 &  [-$\infty, \infty$] & 18.3  & --  \\
& \xmark & 101.7 & 24 & [-$\infty, \infty$] & 17.8  & 11.7 \\ \hline
 & \xmark & 53.5 & 12 & [-40, 40] & 21.0  & 12.9 \\
Tranformer-XL & \xmark & 101.7 & 24 & [-40, 40] & 19.1  & 12.4 \\
 & \cmark & 50.5 & 12 & [-40, 40] & 20.4  & 12.9 \\
  & \cmark & 95.5 & 24 & [-40, 40] &  19.3 & 12.6 \\
 & \cmark & 185.7 & 48 & [-40, 40] &  18.5 & 12.2 \\ \hline
 
\hline
\end{tabular}
\vskip-5mm
\end{table}

\section{Conclusions}
\label{sec:conc}

In this paper, we investigated Transformers for large-scale speech recognition. We presented our approaches to address the issues of scaling up Transformers with cross-machine multi-GPU training, and shown that our approach can train a Transformer with 48 layers and beyond. For streaming Transformers, we discussed the drawbacks of the attention mask based approach, and studied an alternative method based on Transformer-XL, which is much more computationally efficient for inference. We demonstrated that the offline Transformer can outperform BLSTM by 6\% relative with similar number of model parameters, while the streamable Transformer-XL is comparable with LC-BLSTM.

\bibliographystyle{IEEEtran}

\bibliography{bibtex}

\begin{thebibliography}{10}
\providecommand{\url}[1]{#1}
\csname url@samestyle\endcsname
\providecommand{\newblock}{\relax}
\providecommand{\bibinfo}[2]{#2}
\providecommand{\BIBentrySTDinterwordspacing}{\spaceskip=0pt\relax}
\providecommand{\BIBentryALTinterwordstretchfactor}{4}
\providecommand{\BIBentryALTinterwordspacing}{\spaceskip=\fontdimen2\font plus
\BIBentryALTinterwordstretchfactor\fontdimen3\font minus
  \fontdimen4\font\relax}
\providecommand{\BIBforeignlanguage}[2]{{%
\expandafter\ifx\csname l@#1\endcsname\relax
\typeout{** WARNING: IEEEtran.bst: No hyphenation pattern has been}%
\typeout{** loaded for the language `#1'. Using the pattern for}%
\typeout{** the default language instead.}%
\else
\language=\csname l@#1\endcsname
\fi
#2}}
\providecommand{\BIBdecl}{\relax}
\BIBdecl

\bibitem{hochreiter1997long}
S.~Hochreiter and J.~Schmidhuber, ``Long short-term memory,'' \emph{Neural
  computation}, vol.~9, no.~8, pp. 1735--1780, 1997.

\bibitem{vaswani2017attention}
A.~Vaswani, N.~Shazeer, N.~Parmar, J.~Uszkoreit, L.~Jones, A.~N. Gomez,
  {\L}.~Kaiser, and I.~Polosukhin, ``Attention is all you need,'' in
  \emph{Advances in Neural Information Processing Systems}, 2017, pp.
  5998--6008.

\bibitem{dai2018transformer}
Z.~Dai, Z.~Yang, Y.~Yang, W.~W. Cohen, J.~Carbonell, Q.~V. Le, and
  R.~Salakhutdinov, ``Transformer-xl: Language modeling with longer-term
  dependency,'' in \emph{Proc. ICLR}, 2019.

\bibitem{werbos1990backpropagation}
P.~J. Werbos \emph{et~al.}, ``Backpropagation through time: what it does and
  how to do it,'' in \emph{Proceedings of the IEEE}, vol.~78, no.~10, 1990, pp.
  1550--1560.

\bibitem{karita2019comparative}
S.~Karita, N.~Chen, T.~Hayashi, T.~Hori, H.~Inaguma, Z.~Jiang, M.~Someki,
  N.~E.~Y. Soplin, R.~Yamamoto, X.~Wang \emph{et~al.}, ``A comparative study on
  transformer vs rnn in speech applications,'' in \emph{arXiv preprint
  arXiv:1909.06317}, 2019.

\bibitem{dong2018speech}
L.~Dong, S.~Xu, and B.~Xu, ``Speech-transformer: a no-recurrence
  sequence-to-sequence model for speech recognition,'' in \emph{Proc.
  ICASSP}.\hskip 1em plus 0.5em minus 0.4em\relax IEEE, 2018, pp. 5884--5888.

\bibitem{sperber2018self}
M.~Sperber, J.~Niehues, G.~Neubig, S.~St{\"u}ker, and A.~Waibel,
  ``Self-attentional acoustic models,'' in \emph{arXiv preprint
  arXiv:1803.09519}, 018.

\bibitem{tian2019self}
Z.~Tian, J.~Yi, J.~Tao, Y.~Bai, and Z.~Wen, ``Self-attention transducers for
  end-to-end speech recognition,'' in \emph{arXiv preprint arXiv:1909.13037},
  2019.

\bibitem{salazar2019self}
J.~Salazar, K.~Kirchhoff, and Z.~Huang, ``Self-attention networks for
  connectionist temporal classification in speech recognition,'' in \emph{Proc.
  ICASSP}.\hskip 1em plus 0.5em minus 0.4em\relax IEEE, 2019, pp. 7115--7119.

\bibitem{zeyer2019comparison}
A.~Zeyer, P.~Bahar, K.~Irie, R.~Schl{\"u}ter, and H.~Ney, ``A comparison of
  transformer and {LSTM} encoder decoder models for {ASR},'' in \emph{IEEE
  Automatic Speech Recognition and Understanding Workshop, Sentosa, Singapore},
  2019.

\bibitem{han2019state}
K.~J. Han, R.~Prieto, K.~Wu, and T.~Ma, ``State-of-the-art speech recognition
  using multi-stream self-attention with dilated 1d convolutions,'' in
  \emph{arXiv preprint arXiv:1910.00716}, 2019.

\bibitem{wang2019transformer}
Y.~Wang, A.~Mohamed, D.~Le, C.~Liu, A.~Xiao, J.~Mahadeokar, H.~Huang,
  A.~Tjandra, X.~Zhang, F.~Zhang \emph{et~al.}, ``Transformer-based acoustic
  modeling for hybrid speech recognition,'' in \emph{Proc. ICASSP}.\hskip 1em
  plus 0.5em minus 0.4em\relax IEEE, 2020.

\bibitem{moritz2020streaming}
N.~Moritz, T.~Hori, and J.~L. Roux, ``Streaming automatic speech recognition
  with the transformer model,'' \emph{arXiv preprint arXiv:2001.02674}, 2020.

\bibitem{graves2013hybrid}
A.~Graves, N.~Jaitly, and A.-r. Mohamed, ``Hybrid speech recognition with deep
  bidirectional {LSTM},'' in \emph{Proc. ASRU}.\hskip 1em plus 0.5em minus
  0.4em\relax IEEE, 2013, pp. 273--278.

\bibitem{HighwayBLSTM-zhang2016}
Y.~Zhang, G.~Chen, D.~Yu, K.~Yao, S.~Khudanpur, and J.~Glass, ``Highway long
  short-term memory {RNNs} for distant speech recognition,'' in \emph{Proc.
  ICASSP}, 2016.

\bibitem{wang2019semantic}
C.~Wang, Y.~Wu, Y.~Du, J.~Li, S.~Liu, L.~Lu, S.~Ren, G.~Ye, S.~Zhao, and
  M.~Zhou, ``Semantic mask for transformer based end-to-end speech
  recognition,'' \emph{arXiv preprint arXiv:1912.03010}, 2019.

\bibitem{zhang2020transformer}
Q.~Zhang, H.~Lu, H.~Sak, A.~Tripathi, E.~McDermott, S.~Koo, and S.~Kumar,
  ``Transformer transducer: A streamable speech recognition model with
  transformer encoders and {RNN-T} loss,'' in \emph{ICASSP}.\hskip 1em plus
  0.5em minus 0.4em\relax IEEE, 2020, pp. 7829--7833.

\bibitem{povey2018time}
D.~Povey, H.~Hadian, P.~Ghahremani, K.~Li, and S.~Khudanpur, ``A
  time-restricted self-attention layer for {ASR},'' in \emph{Proc.
  ICASSP}.\hskip 1em plus 0.5em minus 0.4em\relax IEEE, 2018, pp. 5874--5878.

\bibitem{bahdanau2014neural}
D.~Bahdanau, K.~Cho, and Y.~Bengio, ``Neural machine translation by jointly
  learning to align and translate,'' in \emph{Proc. ICLR}, 2015.

\bibitem{zhang2019improving}
B.~Zhang, I.~Titov, and R.~Sennrich, ``Improving deep transformer with
  depth-scaled initialization and merged attention,'' in \emph{EMNLP}, 2019.

\bibitem{ba2016layer}
J.~L. Ba, J.~R. Kiros, and G.~E. Hinton, ``Layer normalization,'' in \emph{NIPS
  Deep Learning Symposim}, 2016.

\bibitem{wang2019learning}
Q.~Wang, B.~Li, T.~Xiao, J.~Zhu, C.~Li, D.~F. Wong, and L.~S. Chao, ``Learning
  deep transformer models for machine translation,'' in \emph{ACL}, 2019.

\bibitem{Miao16}
Y.~Miao, J.~Li, Y.~Wang, S.~Zhang, and Y.~Gong, ``Simplifying long short-term
  memory acoustic models for fast training and decoding,'' in \emph{Proc.
  ICASSP}, 2016.

\bibitem{kingma2014adam}
D.~P. Kingma and J.~Ba, ``Adam: A method for stochastic optimization,'' in
  \emph{arXiv preprint arXiv:1412.6980}, 2014.

\bibitem{lu2019transformer}
L.~Lu, ``A transformer with interleaved self-attention and convolution for
  hybrid acoustic models,'' \emph{arXiv preprint arXiv:1910.10352}, 2019.

\bibitem{simonyan2014very}
K.~Simonyan and A.~Zisserman, ``Very deep convolutional networks for
  large-scale image recognition,'' \emph{Proc. ICLR}, 2015.

\bibitem{rae2019compressive}
J.~W. Rae, A.~Potapenko, S.~M. Jayakumar, and T.~P. Lillicrap, ``Compressive
  transformers for long-range sequence modelling,'' \emph{arXiv preprint
  arXiv:1911.05507}, 2019.

\end{thebibliography}

\end{document}